\documentstyle[prl,aps,twocolumn,epsfig]{revtex}
\begin{document}

\draft

\title{$G_{Ep}/G_{Mp}$ ratio by
polarization transfer in $\vec ep\rightarrow e\vec p$}

\author {
M.K.~Jones,$^{1}$
K.A.~Aniol,$^{7}$
F.T.~Baker,$^{4}$
J.~ Berthot,$^{6}$
P.Y.~ Bertin,$^{6}$ 
W.~ Bertozzi,$^{22}$
A.~ Besson,$^{6}$
L.~Bimbot,$^{26}$
W.U.~Boeglin,$^{10}$
E.J.~Brash,$^{5}$
% ...
\vspace{0.1in}
}

\address{
{(The Jefferson Lab Hall A Collaboration)}\\
\vspace{0.1in}
$^{1}$College of William and Mary, Williamsburg, VA 23187\\
% ...
$^{4}$University of Georgia, Athens, GA 30602\\
$^{5}$University of Regina, Regina, SK S4S OA2, Canada\\
$^{6}$Universit\'{e} Blaise Pascal/CNRS-IN2P3, F-63177 Aubi\`{e}re, France\\ 
$^{7}$California State University at Los Angeles, Los Angeles, CA 90032\\
% ...
$^{10}$Florida International University, Miami, FL 33199\\
% ...
$^{22}$Massachusetts Institute of Technology, Cambridge, MA 02139\\
% ...
$^{26}$Institut de Physique Nucl\'{e}aire, F-91406 Orsay, France\\
% ...
}

\date{\today}

\maketitle

\begin{abstract}
The ratio of the proton's elastic electromagnetic 
form factors  $G_{Ep}/G_{Mp}$ was 
obtained by measuring $P_{t}$ 
and $P_{\ell}$, the transverse and longitudinal recoil 
proton polarization, respectively. For elastic 
$\vec e p \rightarrow e\vec p$, $G_{Ep}/G_{Mp}$ is proportional to 
$P_t/P_{\ell}$. 
Simultaneous measurement of $P_{t}$ and $P_{\ell}$ in a polarimeter 
provides good control of the systematic uncertainty. 
The results for the ratio $G_{Ep}/G_{Mp}$ 
show a systematic decrease as $Q^2$ increases
from 0.5 to 3.5 GeV$^2$ , indicating  for the 
first time a definite difference in the spatial
distribution of charge and magnetization currents in the proton.
\end{abstract}

\pacs{25.30.Bf, 13.40.Gp, 24.85.+p}



Understanding  the structure of the nucleon is of fundamental
importance in nuclear and particle physics;
ultimately such an understanding is necessary to describe the strong force.
Certainly, for any QCD based theory, its ability to predict the pion and
nucleon form factors correctly is one of the most stringent test of its
validity, and hence precise data are required.
The electromagnetic interaction provides a unique tool to investigate the
structure of the nucleon. The elastic electromagnetic  form factors  
of the nucleon characterize its internal structure; 
they are connected to its spatial charge and current distributions.

The earliest investigations of the proton form factor by Hofstadter
{\it {et al.}}\cite{hofst} established 
the dominance of the one-photon exchange
process in the elastic $ep$ reaction. 
It indicated that 
the Dirac, $F_{1p}$, and Pauli, $F_{2p}$, form factors  depend 
only on  four-momentum transfer squared which for elastic scattering is
in the space-like region.  $F_{1p}$  and  $F_{2p}$ were found to
have approximately the same $Q^2$ dependence up to
$\approx$ 0.5 GeV$^2$, where $Q^2 = 4E_{e}E^{\prime}_{e}\sin^2\frac{\theta_e}{2}$, $E^{\prime}_{e}$ and $\theta_e$ are the scattered electron's energy and angle and  $E_e$ is the incident beam energy. 
The data were fitted  with a dipole shape, 
$G_D=(1+\frac{Q^2}{0.71})^{-2}$,
characteristic of an exponential radial distribution. 

The elastic $ep$ cross section can be written in terms of the electric,
 $G_{Ep}(Q^2)$, and magnetic, $G_{Mp}(Q^2)$, Sachs form 
factors, which are defined as:
\begin{equation}
 G_{Ep} = F_{1p} - \tau\kappa_p F_{2p} \mbox{~ and~ } G_{Mp}=F_{1p}
+\kappa_p F_{2p},
\label{eq:sach_f1f2}
\end{equation}
where $\tau = Q^2/4M^2$, $\kappa_p$ is the anomalous nucleon magnetic
 moment and $M$ the mass of the proton. 
 In the limit $Q^2\rightarrow~0$, $G_{Ep}=1$ and $G_{Mp}=\mu_p$, the proton 
magnetic moment. 
The unpolarized $ep$ cross section is:
\begin{equation}
\frac{d\sigma }{d\Omega } =  \frac{{\alpha}^2~E^{\prime}_e~\cos^2\frac
{\theta_e}{2}}{4E_e^3~\sin^4\frac{\theta_e}{2}}  
\left[ G_{Ep}^2+\frac{\tau}{\epsilon}G_{Mp}^2 \right]\left (\frac{1}{1+\tau }\right ),
\label{eq:xngegm}
\end{equation} 
where $\epsilon$ is the virtual photon longitudinal polarization, 
$\epsilon=[1+2(1+\tau)\tan^2(\frac{\theta_e}{2})]^{-1}$.

In the Rosenbluth method \cite{rosenbluth}, the 
separation of $G_{Ep}^2$ and $G_{Mp}^2$ is 
achieved  by measuring the cross section at a given $Q^2$ over
a range of $\epsilon$-values that are  obtained by changing the
beam energy and scattered electron angle.   
In Eq.~(\ref{eq:xngegm}) the $G_{Mp}$ part of the cross section is multiplied
by $\tau$; therefore, as $Q^2$ increases, the cross section becomes dominated
by $G_{Mp}$, making the extraction of $G_{Ep}$ more difficult.  Figure 1 shows 
measurements of proton form factors obtained using this method. 
For $Q^2 <$ 1 GeV$^2$ , the uncertainties in both $G_{Ep}$ and $G_{Mp}$ are only a few percent 
and one finds that $G_{Mp}/ \mu_{p} G_D \simeq G_{Ep}/G_D \simeq 1 $. 
For $G_{Ep}$ above $Q^2 = 1$~GeV$^2$, the large uncertainties 
and the divergence in results between different 
experiments, as seen in Fig.~1a, illustrate the difficulties in obtaining 
$G_{Ep}$ by the Rosenbluth method. In contrast, the uncertainties 
on $G_{Mp}$ remain small up to $Q^2$ = 31.2~GeV$^2$\cite{sill}.
\begin{figure}
\epsfig{file=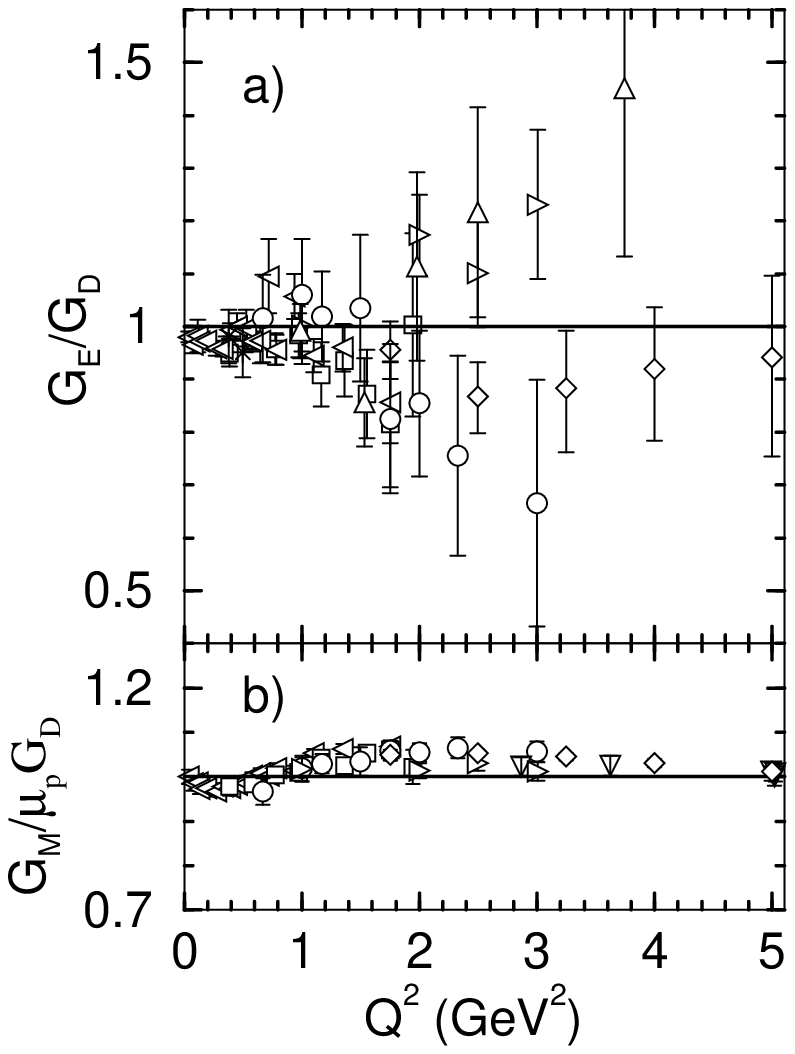}
\caption[]{World data for (a) $G_{Ep}/G_D$ and (b) $G_{Mp}/\mu_p G_D$.
Refs. \cite{litt}$\bigtriangleup$,\cite{berger}$\Box$,\cite{price}$\lhd$,
\cite{bartel}$\circ$,\cite{walker}$\rhd$,
\cite{andivahis}$\diamond$, \cite{milbrath}$\ast$, and \cite{sill}$\bigtriangledown$, versus $Q^2$.}
\label{fig:gepgd_gmpgd} 
\end{figure}

The combination of high energy, 
current and polarization, unique to the Continuous Electron
Beam Accelerator Facility of the 
Jefferson Laboratory (JLab), makes it possible to investigate the
internal structure of the nucleon with higher precision and 
different experimental techniques.
This experiment used the powerful technique of polarization transfer.
 For one-photon
exchange, the scattering of longitudinally polarized electrons results in a
transfer of polarization to the recoil proton with only two 
non-zero 
components, $P_{t}$ perpendicular to, and $P_{\ell }$  parallel to  the proton 
momentum in the scattering plane , given by\cite{akharn}: 
{ 
\begin{eqnarray}
I_{0}P_{t}&=&-2\sqrt{\tau (1+\tau )}G_{Ep}G_{Mp}\tan \frac{\theta_{e}}{2},
\label{eq:pt} \\
I_{0}P_{\ell}&=&\frac{1}{M}(E_{e}+E_{e^{\prime }})\sqrt{\tau (1+\tau )}
G_{Mp}^{2}\tan ^{2}\frac{\theta _{e}}{2},  \label{eq:pl}
\end{eqnarray}
where  $I_{0}=G_{Ep}^{2}+\frac{\tau}{\epsilon} G_{Mp}^{2}$.
Equations~(\ref{eq:pt}) and (\ref{eq:pl}) together give:
\begin{equation}
\frac{G_{Ep}}{G_{Mp}}=-\frac{P_{t}}{P_{\ell}}\frac{(E_{e}+E_{e^{\prime }})} {%
2M}\tan (\frac{\theta _{e}}{2}).
\label{eq:ratio}
\end{equation}
The ratio $G_{Ep}/G_{Mp}$ is obtained from a simultaneous measurement of 
the two recoil polarization components in the polarimeter. Neither the beam polarization nor 
the polarimeter analyzing power needs to be known which results in small 
systematic uncertainties. 
This method was first used recently by Milbrath {\it{et al.}}\cite{milbrath}
at M.I.T.-Bates to measure the ratio $G_{Ep}/G_{Mp}$ 
at low $Q^2$.

Our experiment was done in Hall A at JLab.
Longitudinally polarized electron beams with energies
between 0.934 GeV and 4.090 GeV were scattered in a 15 cm long, 
circulating liquid hydrogen ($ LH_2$) target, refrigerated to 19 K. 
The kinematics settings 
are given in Table I. For the four highest $Q^2$
points, a bulk GaAs photo-cathode excited by circularly polarized laser
light produced beams with $\sim$0.39 polarization and currents up to $\sim$%
115 $\mu$A; the helicity was flipped at 30 Hz. For the lower $Q^2$ 
points, a strained GaAs crystal was used,
and typical polarizations of $\sim$0.60 were achieved with currents up to 
$\sim$15 $\mu$A; the helicity was flipped at 1 Hz. The beam polarization 
was measured with a Mott
polarimeter in the injector line and with a M$\o$ller polarimeter in Hall A. 

Elastic $ep$ events were selected by detecting the scattered
electrons and protons in coincidence in the two identical High Resolution
Spectrometers (HRS) of Hall A\cite{hallA}. 
The HRS deflect particles vertically by 45$^{\circ}$ 
and accept a maximum central trajectory momentum of 4 GeV/c
with a 6.5 msr angular acceptance, $\pm 5\%$ momentum 
acceptance and $\sim 10^{-4}$ momentum resolution. 
The two vertical drift chambers (VDC) installed close to the focal plane 
of each HRS give precise reconstruction of the positions and angles
at the target. 
The trigger was defined by  a coincidence between the 
signals from two scintillator planes in each of the two HRS's.
A focal plane polarimeter (FPP) was installed in the hadron HRS.
In the  FPP, 
two front straw chambers define the
incident proton trajectory and two rear straw
chambers define the proton trajectory after scattering 
in the graphite analyzer\cite{FPP}.  
The graphite analyzer consists of 5
sets of graphite plates that can be moved out to collect 
straight-through trajectories for
alignment of the FPP chambers.   
Graphite thicknesses between 11 to 50~cm were used in order to 
optimize the FPP figure of merit.

The azimuthal angular distribution after a second scattering in
the analyzer of the FPP is given by: 

\begin{eqnarray}
N_{p}(\vartheta ,\varphi )=N_{p}(\vartheta)  [1 &+& (hA_{c}(\vartheta)P_{t}^{fpp}+\alpha)\sin \varphi  \\ \nonumber
 &-& (hA_{c}(\vartheta)P_{n}^{fpp}+\beta)\cos \varphi], \label{eq:phi}
\end{eqnarray}
where $h$ is the electron beam polarization, $N_{p}(\vartheta)$ is the number of 
protons
scattered in the polarimeter, $\vartheta $ and $\varphi $ are the polar and
azimuthal angles after scattering, and $A_{c}(\vartheta )$ is the analyzing 
power; $P_{t}^{fpp}$ and $P_{n}^{fpp}$ are the in-plane 
polarization components, transverse and normal, respectively,
at the FPP analyzer. Instrumental asymmetries ( $\alpha$ and $\beta$ ) 
are canceled by taking 
the difference of the azimuthal distributions for positive and negative electron beam 
helicity. Fourier analysis of this difference distribution 
gives $hA_{c}(\vartheta)P_{t}^{fpp}$ and $hA_{c}(\vartheta)P_{n}^{fpp}$.

The proton spin precesses in the fields of the magnetic elements of the HRS, and therefore
the polarizations at the target and the FPP are different; they are
related through a spin transport matrix 
$\bf P^{fpp}=\bf \left(S\right) \times P$, where 
$\bf P^{fpp}$ and $\bf P$ are polarization column vectors ($n,t,\ell $) at
the FPP and target, respectively, and $\left(\bf S\right)$ is 
the spin transport matrix. A novel method was developed to extract the values of the
polarization components $P_{t}$ and $P_{\ell }$ at the target from the FPP
azimuthal distribution; the integrals in the Fourier analysis  were replaced 
with sums weighted
by the values of the matrix elements, $S_{ij}$, of each event\cite{INP98}. 
The matrix elements $S_{ij}$
depend upon the angular ($\theta \mbox{ and }\phi )$ and
spatial ($y$) coordinates  at the target and proton
momentum ($p$). The  $S_{ij}$'s 
were calculated for each event from the reconstructed
 $y,\phi,\theta,p$ using the spin matrix
determined by a magnetic transport code. Both the
ray-tracing code SNAKE\cite{vernin} and the differential-algebra-based 
code COSY\cite{bertz} were used, and the 
spin precession corrections from both 
methods agree within experimental uncertainties.
The stability of the method was studied in detail for all $Q^2$. The data 
were analyzed in bins of each one of 
the four target variables, one at a time. The results showed that the 
extracted $G_{Ep}/G_{Mp}$ ratio is independent of each of these variables. 
\begin{table}[h]
\caption[]{ The ratio $\mu_p G_{Ep}/G_{Mp} \pm$ statistical uncertainty
(1$\sigma$). $\Delta_{sys}$ is the systematic uncertainty. $\Delta Q^2$ is
half the $Q^2$ acceptance. $<\chi >$ is the average spin precession angle.}
\vspace*{0.15in}
\begin{tabular}{|c|c|c|c|c|}
$<Q^2> \pm \Delta Q^2$ & $E_e$ & $<\chi >$ & $\mu_p G_{Ep}/G_{Mp}$  & $\Delta_{sys}$              \\ \hline
GeV$^2$ & GeV & degrees & $\pm$ stat. uncert.  &     \\ \hline
0.49$\pm$.04 & 0.934 & 105 & 0.966 $\pm$ 0.022 & 0.011   \\ \hline
0.79$\pm$.02 & 0.934 & 118 & 0.950 $\pm$ 0.015 & 0.017   \\ \hline
1.18$\pm$.07 & 1.821 & 136 & 0.869 $\pm$ 0.014 & 0.027   \\ \hline
1.48$\pm$.11 & 3.395 & 150 & 0.798 $\pm$ 0.033 & 0.035   \\ \hline
1.77$\pm$.12 & 3.395 & 164 & 0.728 $\pm$ 0.026 & 0.047   \\ \hline
1.88$\pm$.13 & 4.087 & 168 & 0.720 $\pm$ 0.031 & 0.060        \\ \hline
2.47$\pm$.17 & 4.090 & 196 & 0.726 $\pm$ 0.027 & 0.062   \\ \hline
2.97$\pm$.20 & 4.087 & 218 & 0.612 $\pm$ 0.032 & 0.056        \\ \hline
3.47$\pm$.20 & 4.090 & 239 & 0.609 $\pm$ 0.047 & 0.045   \\
\end{tabular}
\end{table}

The results for the  ratio $\mu_p G_{Ep}/G_{Mp}$ are shown with filled 
circles in Fig.~\ref{fig:gepgmp_prl}a, and as the ratio $Q^2 F_2 /F_1$, obtained from 
Eq.~(\ref{eq:sach_f1f2}), in Fig.~\ref{fig:gepgmp_prl}b; in both figures only the statistical 
uncertainties are plotted as error bars. The data 
are tabulated in Table I, where both statistical  and 
systematic  uncertainties are given for each data point.
Three sources contribute to the systematic 
uncertainty: measurement of the target variables, positioning 
and field strength of the HRS magnetic elements, and uncertainties in the 
dipole fringe-field characterization. The systematic uncertainties 
would shift all data points in the same direction, either up or down. 
No radiative correction 
has been applied to the results. External radiative effects are canceled by 
switching the beam helicity. The internal correction is due to hard
photon emission, two-photon exchange and higher-order contributions. A
dedicated calculation\cite{afanasev} predicts the first
to be of the order of a few per cent. Preliminary indications are that 
the two other contributions are also at the same percentage level.
\begin{figure}[h]
\epsfig{file=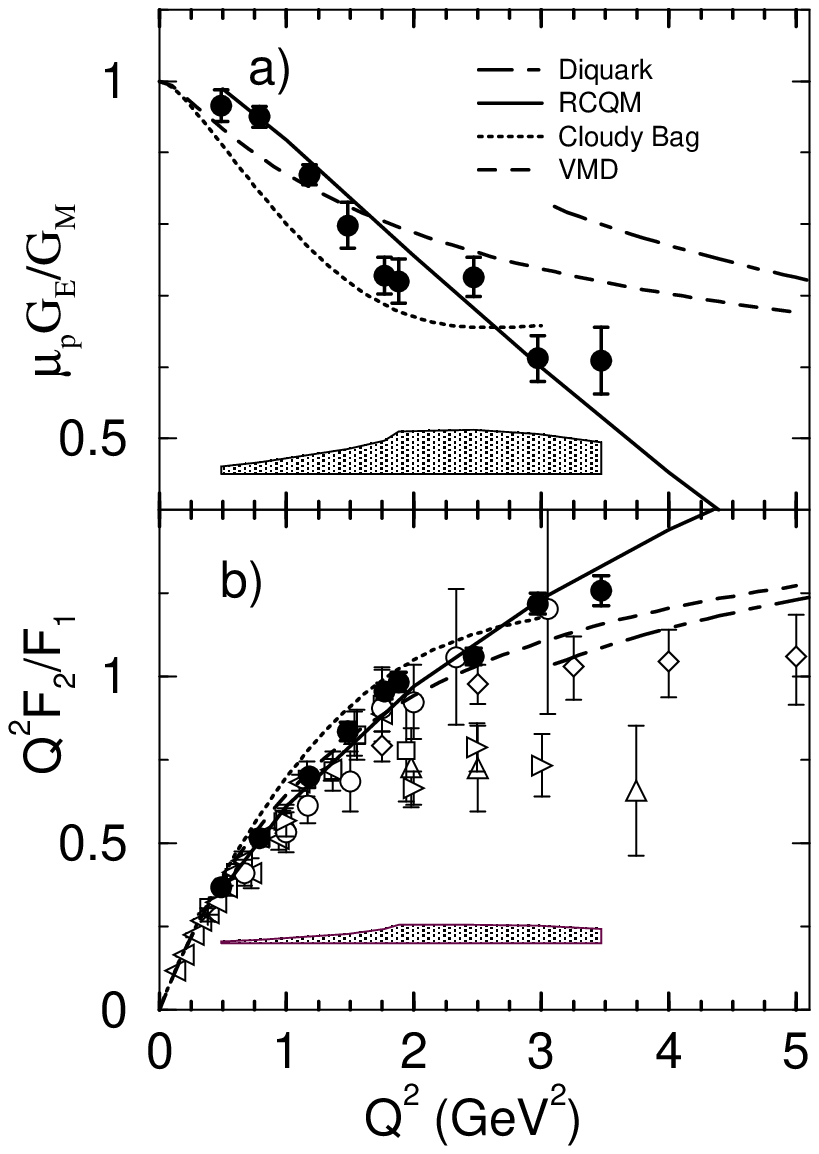}
\caption[]{(a) The ratio $\mu_p G_{Ep}/G_{Mp}$ from this experiment, compared
with theoretical calculations.
(b) The ratio $Q^{2}F_{2p}/F_{1p}$ for the same data, compared to the 
same theoretical models as in (a) and world data; symbols as in Fig.~1.
In both (a) and (b)   
the absolute value of systematic error from this experiment
is shown  by the shaded area. }
\label{fig:gepgmp_prl}
\end{figure}
The most important feature of the data is the sharp decline of the  ratio 
$\mu_{p} G_{Ep}/G_{Mp}$ as $Q^2$ increases, which 
indicates that $G_{Ep}$ falls 
faster than $G_{Mp}$.  Furthermore, as $G_{Mp}/\mu_p G_D$ is approximately 
constant, it follows that $G_{Ep}$ falls more rapidly with $Q^2$ than 
the dipole form factor $G_{D}$. 

Results from this experiment are consistent with the earlier results of refs.
\cite{berger,price,bartel} which have much larger uncertainties. Our results
are compatible with the SLAC data of ref. \cite{andivahis} 
up to about $Q^2$ of 2.5 GeV$^2$, considering the larger uncertainties, 
but our results are in definite 
disagreement with the older results of ref. \cite{walker} from SLAC, as seen 
in Fig.~\ref{fig:gepgmp_prl}b.

The $Q^2 F_2 /F_1$ ratio shown in Fig.~\ref{fig:gepgmp_prl}b indicates a continuing  
increase with $Q^2$, contradicting earlier observations
based on the data of refs. \cite{walker,andivahis} that it might have 
reached a constant value as predicted  in pQCD: $F_1 \sim\frac{1}{Q^4}$ 
and  $F_2 \sim\frac{1}{Q^6}$ \cite{brodsky}. 
It would be of great interest to explore the larger $Q^2$ region where pQCD 
will dominate. Extension of this experiment to larger $Q^2$ has become 
a very interesting prospect and is planned in the near future.

So far,  all 
theoretical models of the nucleon form factors 
are based on effective theories; 
they all rely on a comparison with existing 
data and their parameters are adjustable. 
Much work has been done with the goal of bridging the low and 
high $Q^2$ regimes. There are two quite different approaches 
to calculate nucleon form factors. 
In the first kind, the mesonic degrees of freedom 
are explicit, as in calculations based on Vector Meson 
Dominance (VMD) \cite{hohler,iach,gari,mergell}, models comprising a three-quark core dressed 
with pseudoscalar mesons \cite{dziembowski}, and a calculation based on the solitonic 
nature of the nucleon \cite{holzwarth}. The second kind are QCD-based quark models; these include  
models such as relativistic constituent quark 
(RCQM)\cite{chungcoester,coester,aznau}, 
diquark\cite{kroll}, cloudy bag\cite{lu}, and QCD sum rule\cite{radyushkin}. 
Calculations of the 
nucleon form factors from lattice QCD  are in progress \cite{lattice}.

In Fig.~\ref{fig:gepgmp_prl}, we show as a dashed curve the ratios of  $\mu_p G_{Ep}/G_{Mp}$  
and $Q^{2}F_{2p}/F_{1p}$  calculated from the 
latest published fit to the proton and neutron form factors of Mergell 
{\it{et al.}}\cite{mergell} based on VMD (not including data from this 
experiment). These authors use dispersion relations for the form factors,
with spectral functions taking into account the dominant vector meson poles 
as well as the two-pion channel; an asymptotic behavior 
consistent with pQCD was also included. 

In the earliest study of the RCQM, Chung and 
Coester\cite{chungcoester} investigated 
the effect of the constituent quark masses, the anomalous 
magnetic moment of the quarks, 
$F_{2q}$, and the confinement scale parameter. Recently Coester 
introduced a form factor for $F_{2q}$ 
to reproduce the present data; the result 
is the solid curve in Fig.~\ref{fig:gepgmp_prl}\cite{coester}. 
This illustrates how the
new $G_{Ep}/G_{Mp}$ data can help constrain the 
basic inputs to a particular model.
The dashed-dot curve in Fig.~\ref{fig:gepgmp_prl} shows 
the recently re-evaluated di-quark model prediction of Kroll {\it{et al.}}
\cite{kroll}. In the limit $Q^2\rightarrow \infty$ this model is equivalent 
to the hard-scattering formulation of pQCD. Calculations based 
on the cloudy bag model predict the right slope for $G_{Ep}/G_{Mp}$,
shown as a dotted curve in Fig.~\ref{fig:gepgmp_prl}; this model 
includes an elementary pion field 
coupled to the quarks inside the bag such 
that chiral symmetry is restored\cite{lu}.

Recent theoretical developments indicate that measurements of the  
elastic form factors of the proton to large $Q^2$
may shed light on the problem of nucleon spin. This connection between elastic
form factors and spin has been demonstrated within
the skewed parton distribution (SPD) formalism by Ji \cite{ji}. 
The first moment of the SPD taken in the forward limit yields, 
according to the Angular Momentum Sum 
Rule\cite{ji}, a contribution  to the 
nucleon spin from the quarks and gluons, 
including the orbital angular momentum. 
By subsequently applying the Sum Rule to the SPD, it should become possible 
to estimate the total contribution of the valence quarks to the proton spin
\cite{rad,afan}. 

To conclude, we have presented a new measurement of $G_{Ep}/G_{Mp}$ 
obtained in a polarization transfer experiment with unprecedented accuracy.  
The results demonstrate for the first time that the $Q^2$ dependence 
of $G_{Ep}$ and $G_{Mp}$ is significantly different. The 
quality of the JLab data will place a tight constraint on the theoretical models. 
Results from this experiment combined with future measurements of the neutron 
form factors will bring us closer to a single description of the structure of 
the nucleon. 

The collaboration thanks the Hall A technical staff and the Jefferson Lab 
Accelerator Division for their outstanding support during this experiment. 
This work was supported in part by the U.S. Department of Energy, the U.S. 
National Science Foundation, the 
Italian Istituto Nazionale di Fisica Nucleare (INFN), the French Commissariat 
\`a l'Energie Atomique and Centre National de la Recherche Scientifique 
(CNRS), and the Natural Sciences and Engineering 
Research Council of Canada.


\begin{references}

\bibitem{hofst}  E.E. Chambers and R. Hofstadter, Phys. Rev. {\bf 103}, 1454 (1956).

\bibitem{rosenbluth} M.N. Rosenbluth, Phys. Rev. {\bf 79}, 615 (1950).

\bibitem{litt}   J. Litt {\it et al.}, Phys. Lett. B {\bf 31}, 40 (1970). 

\bibitem{berger}   Ch. Berger {\it et al.}, Phys. Lett. B {\bf 35}, 87 (1971).

\bibitem{price}   L.E. Price {\it et al.}, Phys. Rev. D {\bf 4}, 45 (1971).

\bibitem{bartel}   W. Bartel {\it et al.}, Nuc. Phys. B {\bf 58}, 429 (1973). 
  
\bibitem{walker}   R.C. Walker {\it et al.}, Phys. Rev. D {\bf 49}, 5671 (1994). 

\bibitem{andivahis}   L. Andivahis {\it et al.}, Phys. Rev. D {\bf 50}, 5491 (1994). 

\bibitem{milbrath} B. Milbrath {\it et al.}, Phys. Rev. Lett. {\bf 80}, 452 
(1998); erratum, Phys. Rev. Lett. {\bf 82}, 2221 (1999).

\bibitem{sill} 
 A.F. Sill {\it et al.}, Phys. Rev. D {\bf 48}, 29 (1993).

\bibitem{akharn} A.I. Akhiezer and M.P. Rekalo, Sov. J.
Part. Nucl. {\bf 3}, 277 (1974); R. Arnold, C. Carlson and F. Gross, Phys.
Rev. C {\bf 23}, 363 (1981).

\bibitem{hallA}http://www.jlab.org/Hall-A/equipment/

\bibitem{FPP} M.K. Jones {\it{et al.}}, AIP Conf. Proc. {\bf 412}, 
ed. T.W. Donnelly, p.342 (1997). 
http://www.jlab.org/Hall-A/equipment/detectors/fpp\_overview.html

\bibitem{INP98} G. Qu\'em\'ener {\it{et al.}}, Nucl. Phys. A{\bf 654}, 543c, (1999). 

\bibitem{vernin}P. Vernin, SNAKE, private communication, SPhN-CEA in Saclay, France.

\bibitem{bertz} M. Bertz, COSY INFINITY version 7, NSCL Tech. Rep., MSUCL-977, 
Michigan State University (1995).

\bibitem{afanasev} A.V. Afanasev, I. Akushevich, and N. Merenkov, private 
communication (1999). 

\bibitem{brodsky} S.J. Brodsky and G.R. Farrar, Phys. Rev. D {\bf 11}, 1309 
(1975).

\bibitem{hohler}  G. H\"{o}hler $\it {et~ al.}$, Nucl. Phys. B {\bf 114}, 
 505 (1976).

\bibitem{iach}F. Iachello, A.D. Jackson, and A. Land\'e, Phys. Lett. B {\bf 43}, 
 191 (1973). 
 
\bibitem{gari} M.F. Gari and W. Kr\"{u}mpelmann, Z. Phys. A {\bf 322}, 689 (1985) 
.
\bibitem{mergell} P. Mergell, U.G. Meissner, and  D. Drechsel, Nucl. Phys.
A {\bf 596}, 367 (1996). 

\bibitem{dziembowski} Z. Dziembowski {\it et al.}, Ann. of Phys. {\bf 258}, 1 (1997). 

\bibitem{holzwarth} G. Holzwarth, Z. Phys. A {\bf 356}, 339  (1996).


\bibitem{chungcoester} P.L. Chung and F. Coester, Phys. Rev. D {\bf 44}, 229
(1991).

\bibitem{coester} F. Coester, private communication (1999).

\bibitem{aznau} I.G. Aznauryan, Phys. Lett. B {\bf 316}, 391 (1993).


\bibitem{kroll}   P. Kroll, M. Sch\"{u}rmann, and W. Schweiger, Z. Phys. A 
{\bf 338}, 339 (1991); private communication (1998). 

\bibitem{lu} D.H. Lu, A.W. Thomas, and A.G. Williams, Phys.
Rev. C {\bf 57}, 2628 (1998); D.H. Lu, S.N. Yang and A.W. Thomas, 
preprint ADP 99-36/t373. 

\bibitem{radyushkin} A.V. Radyushkin, Acta Phys. Pol. B {\bf 15}, 40 (1984).

\bibitem{lattice} S. Capitani {\it et al.} ,  Nucl. Phys. B (Proc. Suppl.) {\bf 73}, 294 (1999).

\bibitem{ji} X. Ji, Phys. Rev. D {\bf 55}, (1997) 7114; Phys. Rev. Lett. 
{\bf 78},  610 (1997).

\bibitem{rad} A.V. Radyushkin, JLab-THY-98-10, hep-ph/9803316.

\bibitem{afan} A. V. Afanasev, to be published in Proc. INT/JLab Workshop `Exclusive and Semiexclusive Processes
at High Momentum Transfer', Newport News, VA, May 20-22, 1999, Eds.
C.Carlson, A. Radyushkin,  E-print: hep--ph/9910565.



\end{references}
\end{document}